\documentclass[a4paper,aps,pra,superscriptaddress]{revtex4}

\usepackage{geometry}                
\usepackage{graphicx}
\usepackage{amssymb}
\usepackage{epstopdf}
\usepackage{amsmath}

\usepackage{bm}

\newcommand{\p}{\partial}
\newcommand{\skyrmion}{s}
\newcommand{\Skyrmion}{S}
\newcommand{\polarity}{\lambda}
\newcommand{\half}{\frac{1}{2}}

\begin{document}

\title{Gr\"obli solution for three magnetic vortices}

\author{Stavros Komineas}
\affiliation{Department of Applied Mathematics, University of Crete, 71409 Heraklion, Crete, Greece}
\author{Nikos Papanicolaou}
\affiliation{Department of Physics, University of Crete, 71003 Heraklion, Crete, Greece}
\affiliation{Institute for Theoretical and Computational Physics, University of Crete, Heraklion , Greece}

\begin{abstract}
The dynamics of $N$ point vortices in a fluid is described by the
Helmholtz-Kirchhoff (HK) equations which lead to a completely integrable
Hamiltonian system for $N=2$ or 3 
but chaotic dynamics for $N > 3$.
Here we consider a generalization of the HK equations to describe the  dynamics
of magnetic vortices within a collective-coordinate approximation.
In particular, we analyze in detail the dynamics of a system of three
magnetic vortices by a suitable generalization of the solution for three
point vortices in an ordinary fluid obtained by Gr\"obli more than a century ago.
The significance of our results for the dynamics of ferromagnetic elements
is briefly discussed.
\end{abstract}

\maketitle

\section{Introduction}

Ferromagnetic media are described by the density of magnetic moment or magnetization
$\bm{m}=\bm{m}(\bm{r},t)$ which is a vector field of constant length
that satisfies the highly nonlinear Landau-Lifshitz (LL) equation.
Magnetic bubbles and vortices are special solutions of the LL equation characterized
by a topological invariant sometimes called the Pontryagin index or skyrmion number.
Upon suitable normalization, the skyrmion number is integer for magnetic bubbles
occuring in magnetic films with perpendicular easy-axis anisotropy  \cite{malozemoff}
and half integer for magnetic vortices that may be realized in the case of
magnetic films with easy-plane anistotropy \cite{huber82}.

Most of the early studies were carried out on magnetic films
of infinite extent where vortices and bubbles occur as excited states
above a uniform ground state. More recently it has been
realized that the ground state of a disc-shaped magnetic element,
with a diameter of a few hundred nanometers, is itself a vortex
configuration. Hence the vortex is a nontrivial magnetic
state that can be spontaneously created on magnetic 
elements of finite extent. Such a possibility is of obvious 
significance for device applications and explains the
current interest on this subject \cite{neudert05,waeyenberge06,yamada07}.

The dynamics of magnetic vortices is greatly affected
by the underlying topological structure  \cite{papanicolaou91,komineas96}.
This is described by a topological vorticity
which gives the skyrmion number upon integration over the plane of the magnetic film.
As a result, the essential features of magnetic vortex dynamics
are similar to those displayed by ordinary vortices in fluid
dynamics. A single vortex or antivortex is spontaneously pinned
and thus cannot move freely within the ferromagnetic medium.
However, vortex motion relative to the medium is possible in the
 presence of other vortices, or externally applied magnetic field gradients,
 and displays characteristics analogous to those of two-dimensional
 (2D) motion of electric charges in a uniform magnetic field.
 In particular, two like vortices orbit around each other while
 a vortex-antivortex pair undergoes Kelvin motion. A recent
 review of two-vortex magnetic systems together with
 some progress in the study of three-vortex dynamics may be 
 found in Refs.~\cite{komineas09,komineas08}.

In the present paper we study the dynamics of a system
consisting of $N$ interacting magnetic vortices within a 
collective-coordinate approximation which evades the complexities of the 
complete LL equation but retains important  dynamical features
due to the underlying topology. The relevant equations are 
a generalization of the HK equations encountered in the study 
of $N$ point vortices in ordinary fluids \cite{helmholtz,kirchhoff}. The latter equations
were solved explicitly by Helmholtz himself for the two-vortex
($N=2$) system, while Kirchhoff concluded that the
three-vortex ($N=3$)
system
 is completely integrable. But a complete solution
for $N=3$ was given in the 1877 dissertation of Gr\"obli \cite{groebli}
recently reviewed by Aref et al \cite{aref92} (see also \cite{aref79}). For $N>3$, the system
is not completely integrable and thus leads to chaotic dynamics.

In Section II, we state the appropriate modifications of the 
HK equations to account for $N$  interacting magnetic
vortices, derive the corresponding conservation laws, and
provide a first demonstration with an explicit solution for
two-vortex systems. The generalization of Gr\"obli's solution
to three magnetic vortices is given in Section III and is illustrated
in detail in Section IV for important special cases such as
scattering of a vortex-antivortex pair off
a target vortex initially at rest, three-vortex collapse,
bounded three-vortex motion, etc.
Our main conclusions are summarized in Section V.

\section{Generalization of the Helmholtz-Kirchhoff equations}

We consider a system of $N$ magnetic vortices labeled by
$\alpha=1,2,\ldots N$. Each vortex is characterized by a pair of indices
$(\kappa_\alpha, \polarity_\alpha)$ where $\kappa_\alpha$ is the vortex number
and $\polarity_\alpha$ the polarity,
which take the values $\kappa_\alpha = \pm 1$ and $\polarity_\alpha = \pm 1$ in any combination.
We also define the skyrmion number
\begin{equation}  \label{eq:skyrmionnumber}
\skyrmion_\alpha = \kappa_\alpha\,\polarity_\alpha
\end{equation}
which also takes the distinct values $\skyrmion_\alpha = \pm 1$ (in any combination
for varying $\alpha=1,2,\ldots N$) and differs from the standard definition
of the Pontryagin index by an overall normalization 
($N_\alpha = -\frac{1}{2} \skyrmion_\alpha$) \cite{komineas09,komineas08}.

Now, in a collective-coordinate approximation \cite{thiele73,pokrovskii85,kovalev02}, the
position of each vortex is labeled by a vector $\bm{r}_\alpha=(x_\alpha, y_\alpha)$
in the 2D plane and the energy functional is given by
\begin{equation}  \label{eq:energy}
   E = - \sum_{\alpha < \beta} \kappa_\alpha \kappa_\beta\; \ln|\bm{r}_\alpha-\bm{r}_\beta|
\end{equation}
in suitable units. Note that the energy is a function of
the vortex number $\kappa_\alpha$ but not the polarities $\polarity_\alpha$. The
equations of motion are then written in Hamiltonian form as
\begin{equation}  \label{eq:eqshamilton0}
   \skyrmion_\alpha\,\frac{dx_\alpha}{dt} =  \frac{\p E}{\p y_\alpha},\qquad
   \skyrmion_\alpha\,\frac{dy_\alpha}{dt} = -\frac{\p E}{\p x_\alpha}\,;\quad
   \alpha = 1,2,\ldots N,
\end{equation}
or, more explicitly, as
\begin{equation}  \label{eq:eqshamilton1}
  \polarity_\alpha \frac{dx_\alpha}{dt}  = -\sum_{\beta\neq\alpha} \kappa_\beta \frac{y_\alpha - y_\beta}{|\bm{r}_\alpha-\bm{r}_\beta|^2}, \qquad
  \polarity_\alpha \frac{dy_\alpha}{dt}  = \sum_{\beta\neq\alpha} \kappa_\beta \frac{x_\alpha - x_\beta}{|\bm{r}_\alpha-\bm{r}_\beta|^2}.
\end{equation}
These differ from the HK equations for $N$ point 
vortices in an ordinary fluid by the presence of the
polarities $\polarity_\alpha$ (or the skyrmion numbers $\skyrmion_\alpha$) in addition 
to the vortex numbers $\kappa_\alpha$. Equations \eqref{eq:eqshamilton1} reduce to the
HK equations in the special limit where all 
polarities are taken to be  equal ($\polarity_1= \polarity_2 = \ldots = \polarity_N$). Also
note that the vortex number $\kappa_\alpha$ may assume any continuous
value for a fluid vortex but takes the quantized values $\pm 1$ for
a magnetic vortex.

In addition to the conserved energy of Eq.~\eqref{eq:energy}, the
Hamiltonian system \eqref{eq:eqshamilton0} or \eqref{eq:eqshamilton1} possesses conservation laws
which originate in translational and rotational invariance;
i.e., the analog of linear momentum $\bm{P}= (P_x,P_y)$ with
\begin{equation}  \label{eq:linmom}
P_x = -\sum_{\alpha} \skyrmion_\alpha \; y_\alpha,   \qquad
P_y = \sum_{\alpha} \skyrmion_\alpha \; x_\alpha,
\end{equation}
and angular momentum
\begin{equation}  \label{eq:angmom}
L = \half \sum_{\alpha} \skyrmion_\alpha \; (x^2_\alpha + y^2_\alpha)
\end{equation}
both defined in convenient units. Actually, the physical significance
of these conservation laws depends crucially on whether or not
the total skyrmion number
\begin{equation}  \label{eq:totalskyrmionnumber}
\Skyrmion = \sum_\alpha \skyrmion_\alpha = \sum_\alpha \kappa_\alpha\,\polarity_\alpha
\end{equation}
vanishes. The linear momentum \eqref{eq:linmom} assumes its customary
meaning only when $\Skyrmion=0$. For $\Skyrmion \neq 0$, one may instead 
define the conserved guiding center $\bm{R}=(R_x, R_y)$ with
\begin{equation}  \label{eq:guidingcenter}
R_x = \frac{1}{\Skyrmion}\,\sum_\alpha \skyrmion_\alpha x_\alpha,\qquad
R_y = \frac{1}{\Skyrmion}\,\sum_\alpha \skyrmion_\alpha y_\alpha,
\end{equation}
which is a measure of position rather than linear momentum.
The inherent transmutation of momentum into position occurs
also within the complete LL equation and has been the subject
of several investigations \cite{papanicolaou91,komineas96,komineas08}. To complete the discussion 
of conservation laws we note that the linear momentum 
\eqref{eq:linmom} and the angular momentum \eqref{eq:angmom} may be combined to yield
the derivative conserved quantity
\begin{equation}  \label{eq:angmom2}
\half\,\sum_{\alpha\beta} \skyrmion_\alpha\skyrmion_\beta (\bm{r}_\alpha-\bm{r}_\beta)^2
\end{equation}
which depends only on the relative distances $|\bm{r}_\alpha-\bm{r}_\beta|$ and
will play an important role in the following.

This section is concluded with a simple application
of the general formalism to the case of two-vortex systems
$(N=2)$. Equations \eqref{eq:eqshamilton1} then read
\begin{align}  \label{eq:eqstwovortices0}
  \polarity_1 \frac{dx_1}{dt}  = - \kappa_2 \frac{y_1 - y_2}{d^2},  &   \qquad
  \polarity_2 \frac{dx_2}{dt}  = - \kappa_1 \frac{y_2 - y_1}{d^2},
      \nonumber  \\
  \polarity_1 \frac{dy_1}{dt}  =   \kappa_2 \frac{x_1 - x_2}{d^2} ,  &  \qquad
  \polarity_2 \frac{dy_2}{dt}  =   \kappa_1 \frac{x_2 - x_1}{d^2},
\end{align}
where $d^2 \equiv (x_2-x_1)^2+(y_2-y_1)^2$ is the squared relative distance  between 
the two vortices which is conserved by virtue of the conservation 
of the energy $E$ of Eq.~\eqref{eq:energy} restricted to $N=2$ for any
choice of vortex numbers and polarities. But the details 
of the motion depend on the specific choice of $\kappa$'s and $\polarity$'s
as demonstrated below with two representative examples.

First, we consider a vortex-antivortex pair with
equal polarities:
\begin{equation}  \label{eq:twovortices0}
(\kappa_1, \lambda_1) = (1,1),\qquad (\kappa_2, \lambda_2) = (-1,1),
\end{equation}
hence opposite skyrmion numbers $\skyrmion_1=1,\;\skyrmion_2=-1$ and total
skyrmion number $\Skyrmion=\skyrmion_1+\skyrmion_2=0$. A straightforward solution 
of Eqs.~\eqref{eq:eqstwovortices0} then leads to Kelvin motion where the vortex
and the antivortex move in formation at constant relative
distance $d$, along parallel trajectories that are perpendicular
to the line connecting the vortex and the antivortex, with speed
\begin{equation}  \label{eq:velocitytwovortices0}
V = \frac{1}{d}
\end{equation}
and linear momentum \eqref{eq:linmom} that is parallel to the velocity.

In contrast, a vortex-antivortex pair with opposite
polarities:
\begin{equation}  \label{eq:twovortices1}
(\kappa_1, \lambda_1) = (-1,1),\qquad (\kappa_2, \lambda_2) = (1,-1),
\end{equation}
hence $\skyrmion_1=\skyrmion_2=-1$ and nonvanishing total
skyrmion number $\Skyrmion=\skyrmion_1+\skyrmion_2=-2$, undergoes rotational motion
at constant diameter $d$, around a fixed guiding
center \eqref{eq:guidingcenter} which may be taken to be the origin of the
coordinate system $(x_1+x_2=0=y_1+y_2)$, with angular frequency
\begin{equation}  \label{eq:angfreqtwovortices1}
\omega = \frac{2}{d^2}.
\end{equation}
A detailed study of Kelvin and rotational motion 
was carried out within the complete LL equation in
Refs.~\cite{papanicolaou99} and \cite{komineas07}, respectively, the results of which were
recently reviewed in Ref.~\cite{komineas09}. In particular, their results
confirm the validity of Eqs.~\eqref{eq:velocitytwovortices0} and \eqref{eq:angfreqtwovortices1}
for sufficiently
large relative distance $d$, as expected for the collective-coordinate
approximation adopted in the present paper.

Finally we note the the motion of a two-vortex system is Kelvin-like when the total
skyrmion number vanishes 
($\Skyrmion = \skyrmion_1 + \skyrmion_2 = \kappa_1 \polarity_1 + \kappa_2 \polarity_2 = 0$)
and rotational otherwise
($\Skyrmion = \skyrmion_1 + \skyrmion_2 = \kappa_1 \polarity_1 + \kappa_2 \polarity_2 = \pm 2$)
irrespectively of the detailed configuration of the vortex numbers
$\kappa_1, \kappa_2$ and polarities $\polarity_1, \polarity_2$.

\section{Three magnetic vortices}

Consider a system of three interacting magnetic vortices
with vortex numbers and polarities given by the three pairs
of indices $(\kappa_1, \polarity_1),\,(\kappa_2, \polarity_2),\,(\kappa_3, \polarity_3)$.
The three vortices form  a triangle whose vertices are located at 
$\bm{r}_1=(x_1,y_1),\; \bm{r}_2=(x_2,y_2),\; \bm{r}_3=(x_3,y_3)$ and move about the 2D plane
according to Eqs.~\eqref{eq:eqshamilton1} written here explicitly as
\begin{align}  \label{eq:eqshamilton3v}
   \polarity_1\,\frac{dx_1}{dt} = - \kappa_2 \frac{y_1 - y_2}{C_3^2} + \kappa_3 \frac{y_3 - y_1}{C_2^2},\qquad
   \polarity_1\,\frac{dy_1}{dt} =   \kappa_2 \frac{x_1 - x_2}{C_3^2} - \kappa_3 \frac{x_3 - x_1}{C_2^2},
   \nonumber \\
   \polarity_2\,\frac{dx_2}{dt} = - \kappa_3 \frac{y_2 - y_3}{C_1^2} + \kappa_1 \frac{y_1 - y_2}{C_3^2},\qquad
   \polarity_2\,\frac{dy_2}{dt} =   \kappa_3 \frac{x_2 - x_3}{C_1^2} - \kappa_1 \frac{x_1 - x_2}{C_3^2},
   \\
   \polarity_3\,\frac{dx_3}{dt} = - \kappa_1 \frac{y_3 - y_1}{C_2^2} + \kappa_2 \frac{y_2 - y_3}{C_1^2},\qquad
   \polarity_3\,\frac{dy_3}{dt} =   \kappa_1 \frac{x_3 - x_1}{C_2^2} - \kappa_2 \frac{x_2 - x_3}{C_1^2},
    \nonumber
\end{align}
where
\begin{equation}  \label{eq:c1c2c3}
C_1 = |\bm{r}_2-\bm{r}_3|,\quad C_2 = |\bm{r}_3-\bm{r}_1|,\quad C_3 = |\bm{r}_1-\bm{r}_2|
\end{equation}
are the lengths of the triangle sides.

Our task is to solve the initial-value problem defined
by Eqs.~\eqref{eq:eqshamilton3v} with initial conditions furnished by the $t=0$
configuration of the three vortices. Needless to say, a numerical
solution is straightforward and has indeed been used
to confirm our main analytical results obtained in the following by a
generalization of Gr\"obli's original solution for ordinary
fluid vortices \cite{groebli}.

The key observation is that the scalar quantities $C_1, C_2$ and 
$C_3$ satisfy a closed system of equations of their own, which
may be employed to determine the time evolution of the shape and
size of the vortex triangle, irrespectively of its relative orientation
and motion in the plane. Specifically, as a consequence of Eqs.~\eqref{eq:eqshamilton3v},
\begin{align}  \label{eq:eqsc1c2c3}
\frac{d}{dt}(C_1^2) & = 4 \kappa_1 A \left(\frac{1}{\polarity_3 C_2^2} - \frac{1}{\polarity_2 C_3^2}\right)  \nonumber \\
\frac{d}{dt}(C_2^2) & = 4\kappa_2 A \left(\frac{1}{\polarity_1 C_3^2} - \frac{1}{\polarity_3 C_1^2}\right)  \\
\frac{d}{dt}(C_3^2) & = 4\kappa_3 A \left(\frac{1}{\polarity_2 C_1^2} - \frac{1}{\polarity_1 C_2^2}\right).  \nonumber 
\end{align}
Here $A=\nu |A|$ where $\nu=1$ or $-1$ when the vortices $(1,2,3)$
appear in the counterclockwise or clockwise sense in the plane,
and $|A|$ is the area of the vortex triangle which may be 
expressed entirely in terms of $C_1, C_2$ and $C_3$:
\begin{equation}  \label{eq:trianglearea}
|A| = [C (C-C_1)(C-C_2)(C-C_3)]^{1/2},\qquad C\equiv \half\,(C_1+C_2+C_3).
\end{equation}
Eqs.~\eqref{eq:eqsc1c2c3} constistute
a reduced dynamical system that is completely integrable 
because it possesses two conservation laws; namely
\begin{equation}  \label{eq:energy3v}
   B =   C_1^{1/\kappa_1}\,  C_2^{1/\kappa_2}\, C_3^{1/\kappa_3},\qquad
 \Gamma = \frac{\skyrmion_2 \skyrmion_3 C_1^2 + \skyrmion_3 \skyrmion_1 C_2^2 + \skyrmion_1 \skyrmion_2 C_3^2}{\skyrmion_1+\skyrmion_2+\skyrmion_3},
\end{equation}
where the conservation of $B$ follows from the conservation 
of the energy of Eq.~\eqref{eq:energy}, and $\Gamma$ from the conservation of the
quantity defined in Eq.~\eqref{eq:angmom2}, both restricted to the three-vortex
system and normalized in a manner suitable for subsequent analysis.
The conservation of $B$ and $\Gamma$ may be confirmed directly 
from Eqs.~\eqref{eq:eqsc1c2c3}. Thus, having determined $B$ and $\Gamma$ from the
initial $(t=0)$ values of $C_1, C_2$ and $C_3$, one may use Eqs.~\eqref{eq:energy3v}
to eliminate two of these variables, say, $C_2$ and $C_3$ in favor
of $C_1$ and thereby reduce Eqs.~\eqref{eq:eqsc1c2c3} to a single equation
for $C_1$ that may be integrated by a simple quadrature.

In order to complete the solution we must also determine the relative
orientation and motion of the vortex triangle in the plane
and thus solve for the actual vortex trajectories.
The latter may be
conveniently described in polar coordinates $(r_\alpha, \theta_\alpha),\; \alpha=1,2,3$.
Using these coordinates Hamilton's equations \eqref{eq:eqshamilton0} read
\begin{equation}  \label{eq:eqshamiltonpolar}
   \skyrmion_\alpha\,r_\alpha\,\frac{dr_\alpha}{dt} =  \frac{\p E}{\p \theta_\alpha} \qquad
   \skyrmion_\alpha\,r_\alpha\,\frac{d\theta_\alpha}{dt} = -\frac{\p E}{\p r_\alpha}.
\end{equation}
We write explicitly the three equations for the polar angles:
\begin{align}  \label{eq:eqshamiltonpolarv3}
    \polarity_1\,r_1\frac{d\theta_1}{dt} & = \kappa_2\frac{r_1-r_2\cos(\theta_1-\theta_2)}{C_3^2}
                                                  + \kappa_3\frac{r_1-r_3\cos(\theta_3-\theta_1)}{C_2^2},
   \nonumber \\
   \polarity_2\,r_2\frac{d\theta_2}{dt} & = \kappa_3\frac{r_2-r_3\cos(\theta_2-\theta_3)}{C_1^2}
                                                  + \kappa_1\frac{r_2-r_1\cos(\theta_1-\theta_2)}{C_3^2},  \\
   \polarity_3\,r_3\frac{d\theta_3}{dt} & = \kappa_1\frac{r_3-r_1\cos(\theta_3-\theta_1)}{C_2^2}
                                                  + \kappa_2\frac{r_3-r_2\cos(\theta_2-\theta_3)}{C_1^2}.  
   \nonumber
\end{align}

A crucial step for simplifying the latter equations is to set 
the guiding center \eqref{eq:guidingcenter}
at the origin, i.e., $R_x = 0 = R_y$.
In polar coordinates this constraint reads:
\begin{align}  \label{eq:guidingcenterpolar}
   \skyrmion_1 r_1 \cos\theta_1 + \skyrmion_2 r_2 \cos\theta_2 +  \skyrmion_3 r_3 \cos\theta_3 & = 0 \nonumber  \\
    \skyrmion_1 r_1 \sin\theta_1 + \skyrmion_2 r_2 \sin\theta_2 + \skyrmion_3 r_3 \sin\theta_3  & = 0.
\end{align}
Simple combinations of Eqs.~\eqref{eq:guidingcenterpolar} yield
\begin{align}  \label{eq:cosv3}
2s_2s_3\,r_2r_3\,\cos(\theta_2-\theta_3) & = s_1^2r_1^2 - s_2^2r_2^2-s_3^2r_3^2   \nonumber \\
2s_3s_1\,r_3r_1 \,\cos(\theta_3-\theta_1) & = -s_1^2r_1^2 + s_2^2r_2^2-s_3^2r_3^2  \\
2s_1s_2\,r_1r_2\,\cos(\theta_1-\theta_2) & = -s_1^2r_1^2 - s_2^2r_2^2+s_3^2r_3^2.  \nonumber
\end{align}
Further, we write the sides of the triangle
using polar coordinates:
\begin{align}  \label{eq:c1c2c3cosv3}
C_1^2 & = r_2^2 + r_3^2 - 2r_2r_3\cos(\theta_2-\theta_3) \nonumber \\
C_2^2 & = r_3^2 + r_1^2 - 2r_3r_1\cos(\theta_3-\theta_1)  \\
C_3^2 & = r_1^2 + r_2^2 - 2r_1r_2\cos(\theta_1-\theta_2),  \nonumber
\end{align}
and we can  now use Eqs.~\eqref{eq:cosv3} and \eqref{eq:c1c2c3cosv3} to obtain
explicit relations between the radial coordinates $r_\alpha$ and the
sides of the triangle $C_\alpha$:
\begin{align}  \label{eq:c1c2c3r1r2r3}
\skyrmion_2\skyrmion_3\,C_1^2 & =
    (\skyrmion_2 + \skyrmion_3)\, \Gamma -\skyrmion_1 (\skyrmion_1+\skyrmion_2+\skyrmion_3)\,r_1^2  \nonumber  \\
\skyrmion_3\skyrmion_1\,C_2^2 & = 
    (\skyrmion_3 + \skyrmion_1)\, \Gamma -\skyrmion_2 (\skyrmion_1+\skyrmion_2+\skyrmion_3)\,r_2^2  \\
\skyrmion_1\skyrmion_2\,C_3^2 & =
    (\skyrmion_1 + \skyrmion_2)\, \Gamma -\skyrmion_3 (\skyrmion_1+\skyrmion_2+\skyrmion_3)\,r_3^2,  \nonumber
\end{align}
where $\Gamma$ is the conserved quantity already defined in Eq.~\eqref{eq:energy3v}.

We are now ready to write Eqs.~\eqref{eq:eqshamiltonpolarv3} in a form where
the radial coordinates $r_\alpha$ are eliminated and
only the variables $C_\alpha$ and $\theta_\alpha$ appear.
We use Eqs.~\eqref{eq:c1c2c3cosv3} to eliminate the cosines
and Eqs.~\eqref{eq:c1c2c3r1r2r3} to express the 
radial coordinates in terms of the $C_\alpha$'s
in Eqs.~\eqref{eq:eqshamiltonpolarv3}, to obtain
\begin{align}  \label{eq:angles3v}
2\polarity_1\polarity_2\polarity_3(\skyrmion_1+\skyrmion_2+\skyrmion_3) r_1^2C_2^2C_3^2\,\frac{d\theta_1}{dt} & =
   \polarity_3C_2^2\,[\skyrmion_2\skyrmion_3(-C_1^2+C_2^2+C_3^2) + 2\skyrmion_2^2C_3^2]  \nonumber  \\
  & +\polarity_2C_3^2\,[\skyrmion_2\skyrmion_3(-C_1^2+C_2^2+C_3^2) + 2\skyrmion_3^2C_2^2]   \nonumber \\
2\polarity_1\polarity_2\polarity_3(\skyrmion_1+\skyrmion_2+\skyrmion_3) r_2^2C_3^2C_1^2\,\frac{d\theta_2}{dt} & =
   \polarity_1C_3^2\,[\skyrmion_3\skyrmion_1(C_1^2-C_2^2+C_3^2) + 2\skyrmion_3^2C_1^2]  \\
  & +\polarity_3C_1^2\,[\skyrmion_3\skyrmion_1(C_1^2-C_2^2+C_3^2) + 2\skyrmion_1^2C_3^2]  \nonumber  \\
2\polarity_1\polarity_2\polarity_3(\skyrmion_1+\skyrmion_2+\skyrmion_3) r_3^2C_1^2C_2^2\,\frac{d\theta_3}{dt} & =
   \polarity_2C_1^2\,[\skyrmion_1\skyrmion_2(C_1^2+C_2^2-C_3^2) + 2\skyrmion_1^2C_2^2]  \nonumber  \\
  & +\polarity_1C_2^2\,[\skyrmion_1\skyrmion_2(C_1^2+C_2^2-C_3^2) + 2\skyrmion_2^2C_1^2],\nonumber
\end{align}
where it is understood that the $r_\alpha$'s are given through Eqs.~\eqref{eq:c1c2c3r1r2r3}
in terms of the $C_\alpha$'s.
Explicit applications of the preceding general results are given in Section IV.

\section{An interesting three-vortex system}

For an explicit demonstration we consider a three-vortex
system with vortex numbers and polarities given by
\begin{equation}  \label{eq:threevortices}
(\kappa_1, \lambda_1) = (1,1),\quad
(\kappa_2, \lambda_2) = (-1,1),\quad
(\kappa_3, \lambda_3) = (1,-1)
\end{equation}
and corresponding skyrmion numbers $\skyrmion_1=1,\; \skyrmion_2=-1=\skyrmion_3$.
The pair $(1,2)$ is the vortex -antivortex pair of Eq.~\eqref{eq:twovortices0} which
would undergo Kelvin motion in the absence of vortex 3.
In contrast, the pair $(2,3)$ would undergo rotational motion
in the absence of vortex 1, as discussed in Sec.~II following Eq.~\eqref{eq:twovortices1}.
Thus the motion of the three vortices is expected to be
a combination of Kelvin and rotational motion. Since the
total skyrmion number $\Skyrmion=\skyrmion_1+\skyrmion_2+\skyrmion_3=-1$ is different
from zero, the guiding center for the three-vortex system
defined from Eq.~\eqref{eq:guidingcenter}, i.e.,
\begin{equation}  \label{eq:guidingcenter3v}
R_x = -x_1 + x_2 + x_3,\qquad R_y = -y_1 + y_2 + y_3,
\end{equation}
is conserved and thus remains fixed during the motion.

We now return to the reduced system \eqref{eq:eqsc1c2c3} applied for
the specific choice \eqref{eq:threevortices}:
\begin{align}  \label{eq:eqsc1c2c3v3}
\frac{d}{dt}(C_1^2) & = -4 A \left(\frac{1}{C_2^2} + \frac{1}{C_3^2}\right)  \nonumber \\
\frac{d}{dt}(C_2^2) & = -4 A \left(\frac{1}{C_3^2} + \frac{1}{C_1^2}\right)   \\
\frac{d}{dt}(C_3^2) & = 4 A \left(\frac{1}{C_1^2} - \frac{1}{C_2^2}\right), \nonumber
\end{align}
while the conservation laws defined from Eq.~\eqref{eq:energy3v} are
given by
\begin{equation}  \label{eq:energy3v0}
B = \frac{C_1\,C_3}{C_2},\qquad \Gamma = -C_1^2 + C_2^2 + C_3^2.
\end{equation}
One may then express $C_2$ and $C_3$ in terms of $C_1$ as
\begin{equation}  \label{eq:c2c3}
C_2^2 = \frac{C_1^2+\Gamma}{C_1^2+B^2}\,C_1^2,\qquad 
C_3^2 = \frac{C_1^2+\Gamma}{C_1^2+B^2}\,B^2
\end{equation}
and the area of the triangle defined earlier in Eq.~\eqref{eq:trianglearea} as
\begin{equation}  \label{eq:trianglearea0}
4 A = \nu \sqrt{4 C_2^2 C_3^2 - \Gamma^2} 
= \frac{\nu}{C_1^2+B^2}\,\sqrt{4B^2 C_1^2 (C_1^2+\Gamma)^2-\Gamma^2 (C_1^2+B^2)^2}.
\end{equation}
Then the system of Eqs.~\eqref{eq:eqsc1c2c3v3} reduces to the single equation
\begin{equation}  \label{eq:c1}
\frac{d}{dt} (C_1^2) = -\frac{\nu}{C_1^2+\Gamma}\,\left(\frac{1}{C_1^2}+\frac{1}{B^2}\right)\,
\sqrt{4B^2 C_1^2 (C_1^2+\Gamma)^2-\Gamma^2 (C_1^2+B^2)^2},
\end{equation}
which may readily be integrated to yield $C_1=C_1(t)$, as well
as $C_2=C_2(t)$ and $C_3=C_3(t)$ from Eq.~\eqref{eq:c2c3}, for any
specific values of the conserved quantities $B$ and $\Gamma$ calculated from
the initial three-vortex configuration.

In order to complete the solution we must also determine
the relative orientation and motion of the vortex triangle in the
plane and thus solve for the actual vortex trajectories.
We will follow the derivation given in Section III
and quote only important results that will allow us
to provide a detailed illustration of the three-vortex motion
for the special case of vortex numbers and polarities given
by Eq.~\eqref{eq:threevortices}.

It is convenient to specify a reference frame such that
its origin coincides with the conserved guiding center of
Eq.~\eqref{eq:guidingcenter3v}:
\begin{equation}  \label{eq:guidingcenter3v0}
-x_1 + x_2 + x_3 = 0,\qquad -y_1+y_2+y_3 = 0.
\end{equation}
In this frame the radial distances $r_1, r_2$ and $r_3$ are
related to the triangle sides $C_1, C_2$ and $C_3$ by
Eq.~\eqref{eq:c1c2c3r1r2r3} applied for $\skyrmion_1=1, \skyrmion_2=-1=\skyrmion_3$:
\begin{equation}  \label{eq:r1r2r3}
r_1^2 = 2(C_2^2+C_3^2) - C_1^2,\quad r_2^2=C_2^2,\quad r_3^2=C_3^2,
\end{equation}
while the angular variables satisfy Eqs.~\eqref{eq:angles3v}
now reduced to
\begin{align}  \label{eq:eqshamiltonpolarv30}
\frac{d\theta_1}{dt} & = \frac{\Gamma}{2(C_1^2+2\Gamma)}\left(\frac{1}{C_2^2} - \frac{1}{C_3^2}\right) \nonumber \\
\frac{d\theta_2}{dt} & = -\frac{\Gamma}{2C_2^2}\left(\frac{1}{C_1^2} + \frac{1}{C_3^2}\right) + \frac{1}{C_1^2}   \\
\frac{d\theta_3}{dt} & = \frac{\Gamma}{2C_3^2}\left(\frac{1}{C_2^2} - \frac{1}{C_1^2}\right) + \frac{1}{C_1^2}.  \nonumber
\end{align}

To summarize, for the specific choice of vortex numbers
and polarities given by Eq.~\eqref{eq:threevortices}, the radial variables $r_1, r_2$ and
$r_3$ are expressed in terms of $C_1, C_2$ and $C_3$ from Eqs.~\eqref{eq:r1r2r3}
and ultimately in terms of $C_1$ alone using the conservation
laws as in Eqs.~\eqref{eq:c2c3}; while the time evolution of $C_1$ is 
determined from Eq.~\eqref{eq:c1} by a straightforward integration.
Similarly, we may use Eqs.~\eqref{eq:c2c3} to express the right-hand 
sides of Eq.~\eqref{eq:eqshamiltonpolarv30} in terms of $C_1$ and thus the
time evolution of the angular variables $\theta_1, \theta_2$ and $\theta_3$ is also
reduced to simple quadratures.

\begin{figure}[t]
   \centering
   \includegraphics[width=3.5in]{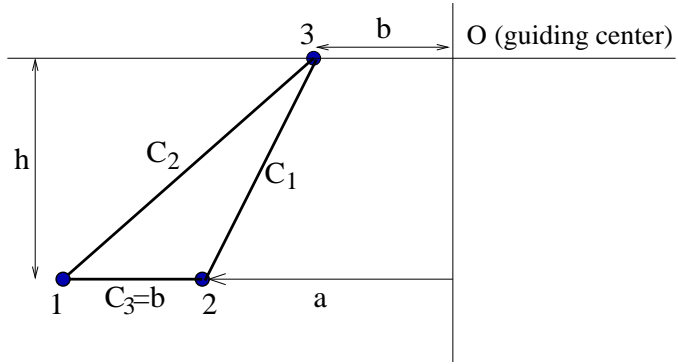} 
   \caption{Initial configuration of three magnetic vortices (1,2,3) with vortex numbers
   and polarities given by Eq.~\eqref{eq:threevortices}. The origin of the coordinate frame
   is taken to coincide with the conserved guiding center of the three-vortex system.
   The impact parameter $a$ is taken to be negative when vortex 2 lies to the left
   of the y-axis and positive otherwise.}
   \label{fig:threevortex}
\end{figure}

Without
loss of generality, the initial three-vortex configuration is
shown in Fig.~\ref{fig:threevortex}. In particular, $b$ is the initial length
of the base $(1,2)$ of the triangle and $h$ the corresponding
height. The impact parameter $a$ is defined as the distance
of vortex 2 from the y-axis and is taken to be negative when
vortex 2 is located to the left of the y-axis and positive
otherwise. Also note that vortex 3 is initially located on the
negative x-axis at a distance $b$ from the origin, thus enforcing 
Eqs.~\eqref{eq:guidingcenter3v0}. Hence, the initial values of the triangle sides
are $C_1^2=h^2+(a+b)^2,\, C_2^2=h^2+a^2$, and $C_3^2=b^2$, and
the conserved quantities of Eq.~\eqref{eq:energy3v0} are given by
\begin{equation}  \label{eq:energyhab}
B^2 = \frac{h^2+(a+b)^2}{h^2+a^2}\,b^2,\qquad \Gamma=-2ab,
\end{equation}
where $\Gamma$ is positive for left impact  ($a<0$) and negative 
for right impact ($a>0$). This completes the description of the
general procedure which is explicitly implemented in the
continuation of this section using examples of increasing complexity.

\subsection{Special case}

%
First we employ the general formalism developed in the
present paper to recover a special solution already given
in the Appendix of our earlier paper \cite{komineas08}. 
The initial
configuration is taken to be an isosceles triangle defined by
choosing the impact parameter $a=-b/2$ in Fig.~\ref{fig:threevortex}.
The vortex-antivortex pair $(1,2)$ would then move upward with initial
velocity $V\approx 1/b$ (Kelvin motion) and eventually collide against
the target vortex 3. After collision the vortex-antivortex pair
moves off to infinity at some scattering angle while the target
vortex comes to rest at a new location. Our aim in the
following is to calculate the scattering angle as well as the final position
of the target vortex.

Now, inserting $a=-b/2$ in Eq.~\eqref{eq:energyhab}, we find that the values
of the conserved quantities are given by
\begin{equation}  \label{eq:energy3v0A}
B^2 = b^2 = \Gamma
\end{equation}
and are independent of the initial triangle height $h$. Then
Eqs.~\eqref{eq:c2c3} read
\begin{equation}  \label{eq:c2c3A}
C_1(t)=C_2(t),\qquad C_3(t) = b
\end{equation}
which establish that the triangle remains isosceles $(C_1=C_2)$
and the length of its base remains constant $(C_3=b)$ at all
times $t>0$. But $C_1$ itself undergoes a nontrivial time
evolution governed by Eq.~\eqref{eq:c1}:
\begin{equation}  \label{eq:c1A}
\frac{d}{dt} (C_1^2) = -\nu b\,\left(\frac{1}{C_1^2}+\frac{1}{b^2}\right)\sqrt{4 C_1^2-1}.
\end{equation}
A more convenient parametrization is obtained by
\begin{equation}
C_1 = \sqrt{H^2+ \frac{1}{4}b^2} = C_2,\qquad C_3 = b,
\end{equation}
where $H=H(t)$ is the instantaneous height of the triangle.
Eq.~\eqref{eq:c1A} then reads
\begin{equation}  \label{eq:H}
\frac{dH}{dt} = -\frac{\nu}{b}\,\frac{H^2+\frac{5}{4}b^2}{H^2+\frac{1}{4}b^2}
\end{equation}
and must be solved with initial condition $H(t=0)=h$.

During the initial stages of the evolution the three vortices $(1,2,3)$
appear in the plane in a counterclockwise sense and thus
$\nu=1$. Eq.~\eqref{eq:H} is then integrated to yield
\begin{align}  \label{eq:Hsolution1}
\frac{H}{\zeta b} - \frac{4}{5} \arctan\left(\frac{H}{\zeta b}\right) = \frac{t_0-t}{\zeta b^2},\qquad
 0 \leq t \leq t_0  \\
 \zeta \equiv \frac{\sqrt{5}}{2},\quad  t_0 \equiv \zeta b^2\,\left[\frac{h}{\zeta b} - - \frac{4}{5} \arctan\left(\frac{h}{\zeta b}\right) \right].  \nonumber
\end{align}
As $t$ approaches $t_0$, height $H$ approaches zero and the triangle
reduces to a straight line element $(1,3,2)$ at $t=t_0$ with
vortex 3 located at the center of the line element $(1,2)$.
Also recalling that the length of the line element $(1,2)$ remains
constant (equal to $b$) we conclude that $t_0$ is the instance of the
closest encounter of the three vortices. For $t > t_0$, vortex 3 emerges
from the other side and thus Eq.~\eqref{eq:H} must be integrated with $\nu=-1$
and initial condition $H(t=t_0)=0$:
\begin{equation}  \label{eq:Hsolution2}
\frac{H}{\zeta b} - \frac{4}{5} \arctan\left(\frac{H}{\zeta b}\right) = \frac{t-t_0}{\zeta b^2},\qquad
  t > t_0.
\end{equation}
In the far future $(t\to\infty)$ $H\to t/b$ which is consistent with a vortex-antivortex pair (1,2)
moving away from the target vortex 3 with asymptotic speed $V=1/b$,
in agreement with Eq.~\eqref{eq:velocitytwovortices0}

The preceding results already suggest a process in which
the vortex-antivortex pair (1,2) initially approaches the target
vortex 3 but eventually moves off to infinity at a 
{\it scattering angle} that remains to be calculated. One must also
ascertain the fate of the target vortex 3 well after collision.
Both of these questions are answered below by applying Eqs.~\eqref{eq:r1r2r3}
and \eqref{eq:eqshamiltonpolarv30} to the special case treated in this subsection.

First, we note that Eqs.~\eqref{eq:r1r2r3} suggest that the three vortices
(1,2,3) together with the origin of the coordinate system $O$
(i.e., the guiding center of the three-vortex system) form
a parallelogram $12O3$ at all times during the collision. In particular,
the third equation \eqref{eq:r1r2r3} establishes that $r_3=C_3=b$ is time
independent, or, vortex 3 moves on a circle whose center coincides with the guiding center
and its radius is a constant $b$ equal
to the constant length of the base of the triangle. Furthermore,
the polar angle $\theta_3$ changes during collision by the same amount as the base of the
triangle (1,2) rotates in the 2D plane.
Therefore, the total change $\Delta\theta_3$ is equal to the scattering angle
of the vortex-antivortex pair (1,2) and also specifies the final
location of the target vortex 3.

\begin{figure}[h]
   \centering
   \includegraphics[width=3in]{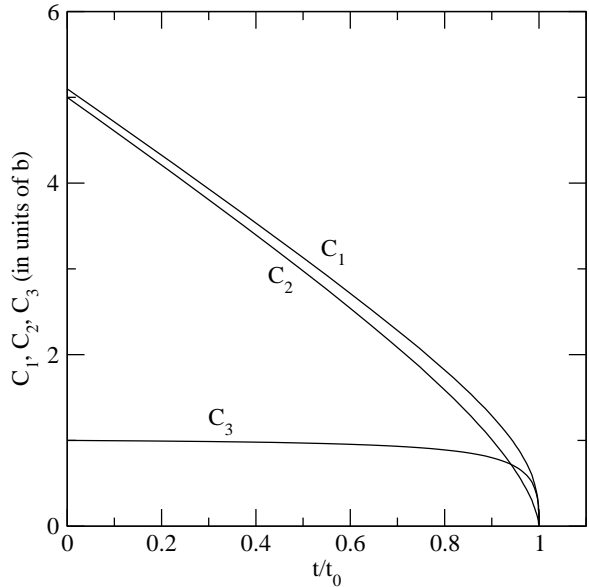} 
   \caption{Time evolution of $C_1, C_2$ and $C_3$ for an initial three-vortex triangle
   given by Fig.~\ref{fig:threevortex} with $a=0, b=1, h=5$. Note that all three lengths
   simultaneously collapse to zero at a finite time $t_0$ calculated from
   Eq.~\eqref{eq:t0}.}
   \label{fig:2}
\end{figure}

In order to actually calculate $\Delta\theta_3$ in the present special
case, we apply Eq.~\eqref{eq:eqshamiltonpolarv30} with $C_1=C_2$:
\begin{equation}  \label{eq:theta3A}
\frac{d\theta_3}{dt} = \frac{1}{C_1^2} = \frac{1}{H^2+\frac{1}{4}b^2},
\end{equation}
where we may insert $H=H(t)$ from Eqs.~\eqref{eq:Hsolution1} and \eqref{eq:Hsolution2}
to obtain $\theta_3=\theta_3(t)$ by a further integration. In fact, a more direct
calculation of the total scattering angle $\Delta\theta_3$ is obtained
by combining Eqs.~\eqref{eq:H} and \eqref{eq:theta3A} to write
\begin{equation}
\frac{d\theta_3}{dH} = -\frac{b}{\nu}\,\frac{1}{H^2+\frac{5}{4}b^2},
\end{equation}
and
\begin{equation}
\Delta\theta_3 = b\int_0^h \frac{dH}{H^2+\frac{5}{4}b^2} + b\int_0^\infty \frac{dH}{H^2+\frac{5}{4}b^2},
\end{equation}
where the two terms may be interpreted as the scattering angles
before and after collision and correspond to the two branches of
the solution $H=H(t)$ given by Eqs.~\eqref{eq:Hsolution1} and \eqref{eq:Hsolution2}.
In particular, for pure scattering where the vortex-antivortex pair originates very far
from the target vortex ($h\to\infty$):
\begin{equation}  \label{eq:2pisqrt5}
\Delta\theta_3 = 2b \int_0^\infty \frac{dH}{H^2+\frac{5}{4}b^2}
 = \frac{2\pi}{\sqrt{5}},
\end{equation}
which reproduces the special result given in the Appendix of Ref.~\cite{komineas08}.
A detailed illustration of the actual vortex trajectories may be found in Fig.~A1
of the above reference (using a different convention for the origin of the coordinate system)
and in Fig.~\ref{fig:5} of the present paper
(where the origin of the coordinate system coincides with the guiding center).

\subsection{Three-vortex collapse}

At first sight, the special example treated in the preceding 
subsection would seem to correspond to a head-on collision
of a vortex-antivortex pair off a target vortex. However,
such an interpretation seems to be unjustified in view of the
peculiar scattering angle calculated in Eq.~\eqref{eq:2pisqrt5}. Indeed, an
example that is closer to head-on collision is obtained 
by choosing vanishing impact parameter $(a=0)$ in the
initial vortex configuration shown in Fig.~\ref{fig:threevortex}.

The initial vortex triangle is then orthogonal, with initial
values of the length of its sides $C_1(t=0) = \sqrt{h^2+b^2},\, C_2 (t=0) = h$
and $C_3 (t=0) = b$. The two conserved quantities of Eq.~\eqref{eq:energy3v0}
now read
\begin{equation}  \label{eq:energy3v0B}
B = \frac{C_1 C_3}{C_2} = \frac{b}{h}\,\sqrt{h^2+b^2},\quad
\Gamma = -C_1^2 + C_2^2 + C_3^2 = 0.
\end{equation}
A notable consequence of the conservation of $\Gamma=0$ is that
the vortex triangle remains orthogonal at all times, even though
the length of its sides $C_1, C_2$, and $C_3$, as well as its relative
orientation in the 2D plane, undergo a nontrivial time
evolution.

In particular, the evolution of $C_1$ is governed by Eq.~\eqref{eq:c1}
applied with $\Gamma=0$ and $B$ given by Eq.~\eqref{eq:energy3v0B}:
\begin{equation}  \label{eq:c1B}
\frac{dC_1}{dt} = - \frac{C_1^2+B^2}{B\, C_1^2},
\end{equation}
while $C_2$, and $C_3$ obtained from Eq.~\eqref{eq:c2c3} are now given by
\begin{equation}  \label{eq:c2c3B}
C_2 = \frac{C_1^2}{\sqrt{C_1^2+B^2}},\quad 
C_3 = \frac{B\,C_1}{\sqrt{C_1^2+B^2}}.
\end{equation}
Eq.~\eqref{eq:c1B} may easily be integrated to yield
\begin{equation}  \label{eq:collapsesolution}
\frac{C_1}{B} - \arctan\left(\frac{C_1}{B}\right) = \frac{t_0-t}{B^2},
\end{equation}
where the integration constant $t_0$ is calculated from the
initial condition $C_1 (t=0) = \sqrt{h^2+b^2}$, or
\begin{equation}  \label{eq:t0}
t_0 = \frac{b^2}{h^2}\,(h^2+b^2)\,\left[ \frac{h}{b} - \arctan\left(\frac{h}{b}\right)\right].
\end{equation}
The physical significance of $t_0$ becomes apparent by applying
Eq.~\eqref{eq:collapsesolution} in the limit $t \to t_0$:
\begin{equation}  \label{eq:c1t0}
C_1 (t\to t_0) \sim [3B (t_0-t)]^{1/3},
\end{equation}
and Eqs.~\eqref{eq:c2c3B} in the same limit to yield
\begin{equation}  \label{eq:c2c3t0}
C_2 (t\to t_0) \sim \frac{1}{B}\,[3B (t_0-t)]^{2/3},\quad
C_3 (t\to t_0) \sim [3B (t_0-t)]^{1/3}.
\end{equation}
Therefore, $t_0$ calculated explicitly from Eq.~\eqref{eq:t0} is the
instance at which all three $C_1, C_2$ and $C_3$ vanish simultaneously,
or, the vortex triangle {\it collapses} to a point. Also taking
into account Eq.~\eqref{eq:r1r2r3}, we conclude that the vortex triangle
collapses to the guiding center $(r_1 = r_2 = r_3 = 0)$ of the three-vortex
system in a finite time interval $t_0$. The exact time dependence
of $C_1, C_2$ and $C_3$ calculated from Eqs.~\eqref{eq:collapsesolution}
and \eqref{eq:c2c3B} over the
entire time interval $0 \leq t \leq t_0$ is depicted in Fig.~\ref{fig:2}.

\begin{figure}[h]
   \centering
   \includegraphics[width=3in]{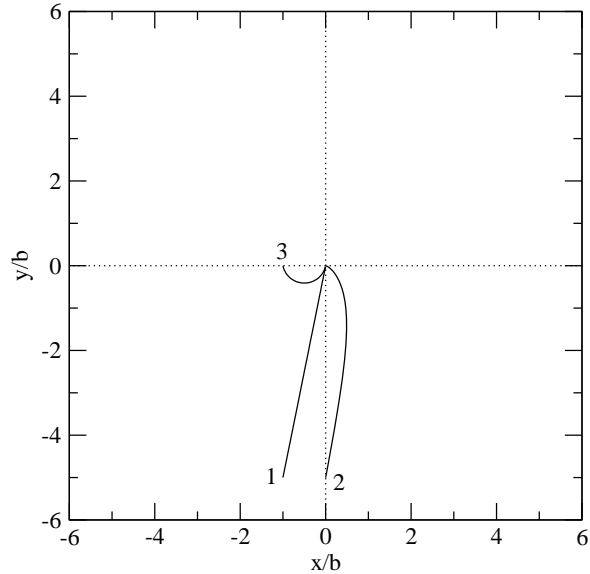} 
   \caption{Trajectories of the three vortices 1,2 and 3 emanating form an initial
   three-vortex configuration given by Fig.~\ref{fig:threevortex} with $a=0, b=1, h=5$.
   Note that the three trajectories merge at the origin of the coordinate frame
   which is taken to coincide with the guiding center of the three-vortex system.}
   \label{fig:3}
\end{figure}

Complete information about the vortex trajectories may be 
obtained by further utilizing the results of Eqs.~\eqref{eq:r1r2r3} 
and \eqref{eq:eqshamiltonpolarv30} or by
a direct numerical integration of Eqs.~\eqref{eq:eqshamilton3v}. The actual trajectories
are illustrated in Fig.~\ref{fig:3} which explicitly demonstrates the
predicted three-vortex collapse. A notable feature of Fig.~\ref{fig:3}
is that vortex 1 moves on a straight line that connects
its original position with the guiding center of the three-vortex system.

Although this special case of vanishing impact parameter
$(a=0)$ is clearly singular from the point of view of scattering 
theory, it is still possible to calculate the scattering angle
$\Delta\theta_3$ before collapse. First, we apply the third of 
Eqs.~\eqref{eq:eqshamiltonpolarv30} with $\Gamma=0$:
\begin{equation}
\frac{d\theta_3}{dt} = \frac{1}{C_1^2}
\end{equation}
and combine this result with Eq.~\eqref{eq:c1B} to obtain
\begin{equation}  \label{eq:dtheta3B}
d\theta_3 = -\frac{B\, dC_1}{B^2+C_1^2}.
\end{equation}
The scattering angle before collapse is then obtained by integrating
Eq.~\eqref{eq:dtheta3B} between the values $C_1(t=0) = \sqrt{h^2+b^2}$ and $C_1(t=0) =0$:
\begin{equation}  \label{eq:}
\Delta\theta_3|_{\rm before} = \int_0^{\sqrt{h^2+b^2}} \frac{B\, dC_1}{B^2+C_1^2}
= \arctan\left( \frac{h}{b}\right).
\end{equation}
In the limit where the vortex-antivortex pair originates
at a large distance $h\to\infty$ from the target vortex,
we find $\Delta\theta_3|_{\rm before} = \pi/2$ and, by convention,
$\Delta\theta_3|_{\rm total} =2(\pi/2) = \pi$.
This result agrees with the interpretation of this special
case as a head-on collision whose singular nature will also
become apparent in the discussion of the following subsection.

\subsection{Scattering in the general case}
Our aim in this subsection is to determine the scattering
angle as a function of impact parameter $a$. The general
strategy is suggested by the special examples treated in the
preceding two subsections. Thus we return to the initial
configuration depicted in Fig.~\ref{fig:threevortex}, now applied for arbitraty
$a$, which yields values for the conserved quantities $B$ and
$\Gamma$ as given in Eq.~\eqref{eq:energyhab}. In particular, we shall be interested
in the limit $h\to \infty$:
\begin{equation}
B^2 = b^2,\quad \Gamma = -2 a b,
\end{equation}
which corresponds to pure (asymptotic) scattering where the
vortex-antivortex pair is initially located very far from
the target vortex and also diverges to infinity after collision.

In order to calculate the scattering angle we return
to Eq.~\eqref{eq:eqshamiltonpolarv30} and express its right-hand side entirely in terms
of $C_1$ using Eq.~\eqref{eq:c2c3} to eliminate $C_2$ and $C_3$. We then combine
the resulting equation with Eq.~\eqref{eq:c1} to write
\begin{equation}  \label{eq:theta3B}
\frac{d\theta_3}{d C_1^2} = -\frac{1}{\nu}\,
\frac{\frac{\Gamma}{2} \frac{B^2-\Gamma}{C_1^2+\Gamma} + \frac{C_1^2+\Gamma}{C_1^2+B^2}\,B^2}
        {\sqrt{4B^2 C_1^2 (C_1^2+\Gamma)^2 - \Gamma^2 (C_1^2+B^2)^2}}.
\end{equation}
Simple inspection of this equation suggests using scaled
variables $u$ and $\gamma$ from
\begin{equation}
u \equiv \frac{C_1^2}{B^2},\qquad \gamma \equiv \frac{\Gamma}{B^2} = -\frac{2a}{b}
\end{equation}
to obtain
\begin{equation}  \label{eq:Dtheta3DuC}
\frac{d\theta_3}{d u} = -\frac{1}{\nu}\,
\frac{\frac{\gamma (1-\gamma)}{2(u+ \gamma)} + \frac{u+ \gamma}{u+1}}
        {\sqrt{4u (u+ \gamma)^2 - \gamma^2 (u+1)^2}}.
\end{equation}
Based on this equation the total scattering angle is calculated
as a function of the parameter $\gamma$ alone by a procedure similar
to that employed in the special example treated in subsection A:
\begin{equation}  \label{eq:Dtheta3C}
\Delta\theta_3 = 2 \int_{u_0}^\infty
\frac{\frac{\gamma (1-\gamma)}{2(u+ \gamma)} + \frac{u+ \gamma}{u+1}}
        {\sqrt{4u (u+ \gamma)^2 - \gamma^2 (u+1)^2}}\, du
\end{equation}
where the factor of two and the upper limit of integration are dictated by the fact
that the vortex-antivortex pair is located very far from the
target vortex well before and after the collision. The lower
limit $u_0$ is a specific value of $u=C_1^2/B^2$ for which the
area of the vortex triangle vanishes. Therefore, $u_0$ must be
chosen among the three roots of the cubic equation
\begin{equation}  \label{eq:cubicequation}
4u(u+\gamma)^2 - \gamma^2 (u+1)^2 = 0.
\end{equation}
It is thus important to examine the behavior of the roots
as functions of the parameter $\gamma$.

For $\gamma < 0$ there exists a real and positive root $u_1$ and
two complex roots $u_2$ and $u_3$ such that $u_2=u_3^*$. In this region
the lower limit in the integral of Eq.~\eqref{eq:Dtheta3C} must be chosen as
$u_0=u_1$. Now, $\gamma=0$ is an exceptional point in that all
three roots vanish $(u_1=u_2=u_3=0)$ and thus the lower
limit must be chosen as $u_0=0$. Applying Eq.~\eqref{eq:Dtheta3C}
for $\gamma=0$ and $u_0=0$ yields
\begin{equation}
\Delta\theta_3 =\int_0^\infty \frac{du}{(u+1) \sqrt{u}} = \pi
\end{equation}
in agreement with the conclusion of Subsection B. Indeed, 
$\gamma=0$ corresponds to vanishing impact parameter $(a=0)$  which
is the special case discussed in Subsection B (three-vortex collapse).

Next we consider the region $0 < \gamma < 1$ where the cubic equation 
again possesses a real and positive root $u_1$ and two complex
roots $u_2=u_3^*$. Thus the lower limit must again be chosen
as $u_0=u_1$. Now, $\gamma=1$ is another exceptional point
in that all three roots become real $(u_1=\frac{1}{4}, u_2=-1=u_3)$
and the lower limit must be chosen to coincide with the
positive root $u_0=u_1=\frac{1}{4}$. Applying Eq.~\eqref{eq:Dtheta3C} for $\gamma=1$ and
$u_0=\frac{1}{4}$ yields
\begin{equation}
\Delta\theta_3 =2\int_{\frac{1}{4}}^\infty \frac{du}{(u+1) \sqrt{4u-1}} = \frac{2\pi}{\sqrt{5}},
\end{equation}
which coincides with the result of Eq.~\eqref{eq:2pisqrt5}.
This is not surprising because $\gamma=-2a/b=1$ leads to an impact parameter
$a=-b/2$ which is indeed the special case discussed in Subsection A.

For $\gamma > 1$, the cubic equation again possesses a real
and positive root $u_1$ and two complex roots $u_2=u_3^*$, thus
we must choose $u_0=u_1$. This generic picture continues to hold
until $\gamma$ reaches the critical value
\begin{equation}  \label{eq:gammac}
\gamma_c = \frac{27}{2}
\end{equation}
where all three roots become real $(u_1=0.5625, u_2=9=u_3)$
and remain real and positive for $\gamma > \gamma_c$. In this range
of $\gamma$, the three roots may be ordered
as $u_1 < u_2 < u_3$ and the lower limit of integration
in Eq.~\eqref{eq:Dtheta3C} must now be chosen as $u_0 = u_3$.

\begin{figure}[h]
   \centering
   \includegraphics[width=3.5in]{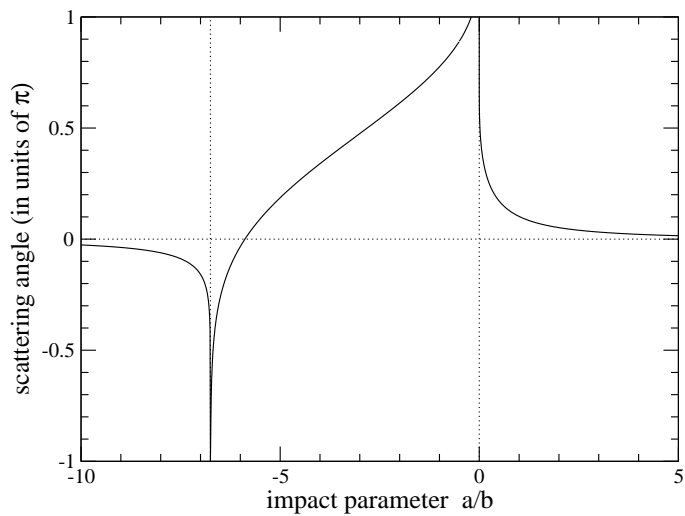} 
   \caption{Scattering angle during collision of a vortex-antivortex (Kelvin) pair (1,2)
   with a target vortex 3, as a function of impact parameter $a$.
   Note the crossover behavior at the critical values $a/b=0$ and $a/b=-27/4$
   discussed in the text. The results depicted in this figure should be read
   modulo $2\pi$.}
   \label{fig:4}
\end{figure}

In all cases the denominator in Eq.~\eqref{eq:Dtheta3C} does not possess
singularities in the domain of integration $[u_0, \infty)$ except for
an integrable square root singularity at the lower end. It can also be shown that
the numerator in Eq.~\eqref{eq:Dtheta3C} does not exhibit
singularities in the integration domain. Thus we have provided
a complete prescription for the calculation of the scattering
angle, which entails locating the appropriate root $u_0=u_0(\gamma)$
of the cubic equation and an elementary numerical integration
of the integral in Eq.~\eqref{eq:Dtheta3C} for any value of $\gamma=-2a/b$.
The calculated scattering angle as a function of impact
parameter $a$ measured in units of $b$ is depicted in Fig.~\ref{fig:4}.

Thus the scattering angle displays critical (crossover)
behavior at two characteristic values of the impact parameter;
namely, $a/b=0$, which corresponds to the three-vortex
collapse discussed in subsection B, and $a/b=-27/4$, which
corresponds to the critical coupling $\gamma=-2a/b=27/2$ of Eq.~\eqref{eq:gammac}.

\begin{figure}[h]
   \centering
   \includegraphics[width=4in]{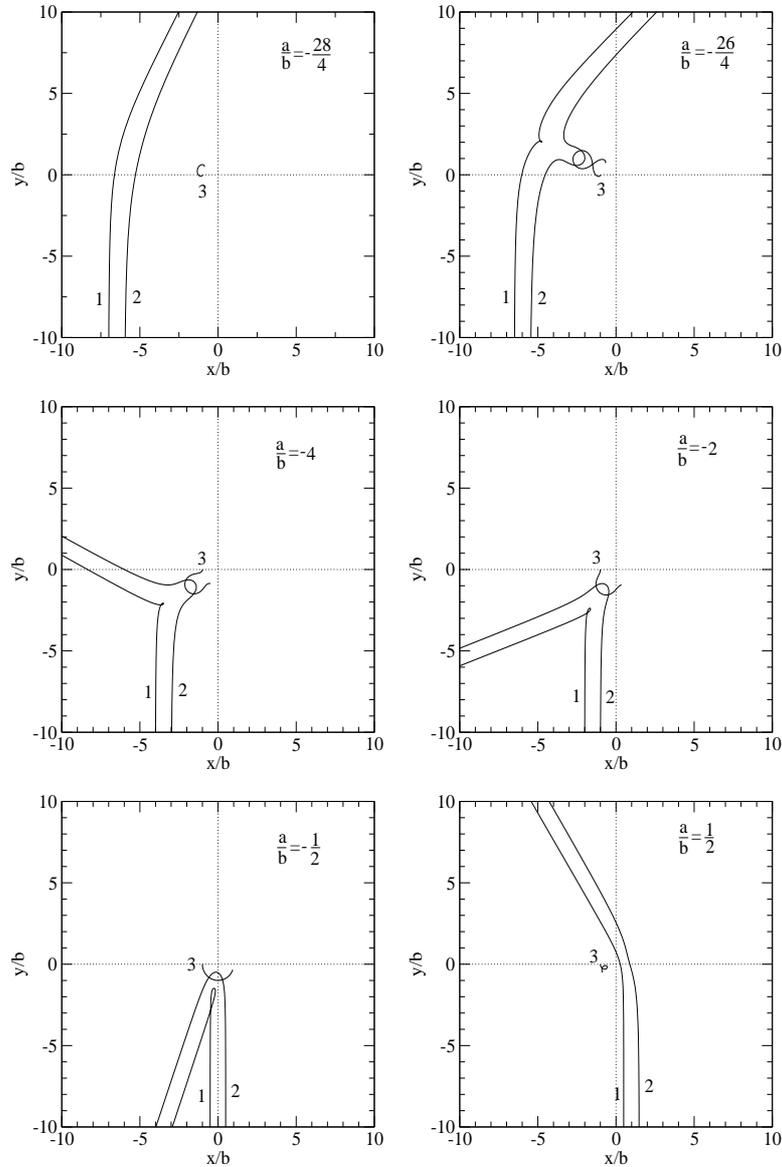} 
   \caption{Vortex trajectories for six characteristic values of the impact parameter $a$.
   In all cases a vortex-antivortex (Kelvin) pair (1,2) collides with a target vortex 3 and is
   deflected by a scattering angle in agreement with the results of Fig.~\ref{fig:4}.
   The target vortex 3 is at rest well before collision and also comes to rest well
   after collision, as expected for an isolated vortex which is always spontaneously pinned.}
   \label{fig:5}
\end{figure}

In Fig.~\ref{fig:5} we display the actual vortex trajectories
calculated numerically from Eqs.~\eqref{eq:eqshamilton3v} for a set of six
characteristic values of $a/b$. The change of pattern becomes
clear as one crosses the two critical values $a/b=0$ and
$a/b=-27/4$. The nature of the critical parameter $\gamma_c$ of Eq.~\eqref{eq:gammac}
is further illuminated in the following subsection.

\subsection{Bounded three-vortex motion}

The analysis of the cubic equation \eqref{eq:cubicequation} revealed
the importance of the critical parameter $\gamma_c=27/2$ of
Eq.~\eqref{eq:gammac} above which all three roots are real (and positive).
We may then order the roots as $u_1 < u_2 < u_3$ and note that
the argument under the square root in, say, Eq.~\eqref{eq:Dtheta3DuC} is
positive when either $u > u_3$ or $u_1 < u < u_2$. The former
case,
\begin{equation}  \label{eq:case1D}
\gamma > \gamma_c = \frac{27}{2},\qquad u > u_3,
\end{equation}
corresponds to the scattering solution discussed in subsection C
where $u=C_1^2/B^2$ approaches infinity well before or after
the collision, while it reaches its minimum value $u_3$ at some
instance during collision. In contrast, the case
\begin{equation}  \label{eq:case2D}
\gamma > \gamma_c = \frac{27}{2},\qquad u_1 < u < u_2,
\end{equation}
may lead to a motion where $u$ oscillates between the values
$u_1$ and $u_2$. Also taking into account Eq.~\eqref{eq:c2c3} we conclude
that all three $C_1, C_2$ and $C_3$ would then remain bounded.
The aim of this subsection is to explicitly construct a
special class of solutions that realize Eqs.~\eqref{eq:case2D} and
thus lead to bounded (quasiperiodic) three-vortex motion.
In order to achieve the two conditions displayed in
Eqs.~\eqref{eq:case2D} we consider an initial configuration where the
three vortices (1,2,3) lie on a straight line in the order
indicated and thus the relative distances are such that $C_1+C_3=C_2$. Specifically, we
choose
\begin{equation}  \label{eq:epsilon}
C_1 = (2 + \epsilon) b,\quad C_2 = (3 + \epsilon) b,\quad C_3=b
\end{equation}
and initially place the three vortices along the negative x-axis
at points such that
\begin{align}  \label{eq:guidingcenterD}
& x_1 = -C_2-C_3,\quad x_2 = -C_2,\quad x_3= -C_3,  \nonumber \\
& y_1 = y_2 = y_3 = 0.
\end{align}
Then the origin of the coordinate system coincides with the
guiding center defined by Eqs.~\eqref{eq:guidingcenter3v}.
Our aim is to calculate the vortex trajectories emanating from the initial configuration
defined by Eqs.~\eqref{eq:epsilon} and \eqref{eq:guidingcenterD}.

We first consider the two conserved quantities calculated from such an
initial configuration,
\begin{equation}
B^2 = \frac{C_1^2 C_3^2}{C_2^2} = \left(\frac{2+\epsilon}{3+\epsilon}\right)^2\,b^2,\quad
\Gamma = -C_1^2 + C_2^2 + C_3^2 = 2(3+\epsilon) b^2,
\end{equation}
and thus
\begin{equation}
\gamma = \frac{\Gamma}{B^2} = \frac{2(3+ \epsilon)^3}{(2+ \epsilon)^2}
\end{equation}
is now the important dimensionless quantity that controls
the behavior of the roots of the cubic equation. Note that
$\epsilon=0$ leads to the critical parameter $\gamma_c=27/2$
of Eq.~\eqref{eq:gammac} while, interestingly, $\gamma > \gamma_c$ for both positive
and negative values of the parameter $\epsilon$.

\begin{figure}[h]
   \includegraphics[width=3in]{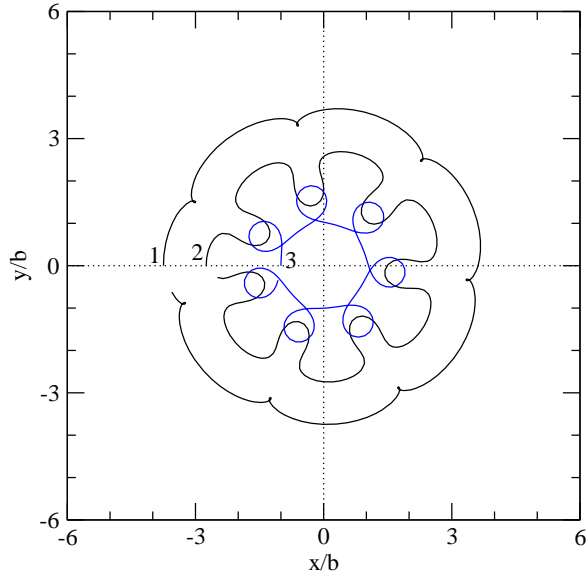} 
   \caption{Trajectories of three vortices in bounded (quasiperiodic)
   motion emanating from an initial configuration specified by
   Eqs.~\eqref{eq:epsilon} and \eqref{eq:guidingcenterD} with $\epsilon=-\frac{1}{4}$.}
   \label{fig:6}
\end{figure}

A numerical solution of the initial value problem
reveals an interesting picture. For $\epsilon > 0$, we encouter
a situation of the type described by Eq.~\eqref{eq:case1D} which
corresponds to a scattering solution where the vortex-antivortex
pair (1,2) moves off to infinity leaving behind vortex 3
which relocates to a new final position. However, $\epsilon < 0$
leads to a situation of the type anticipated by Eq.~\eqref{eq:case2D}.
The resulting bounded (quasiperiodic) three-vortex motion
is illustrated in Fig.~\ref{fig:6} for $\epsilon=-1/4$.
Although the present calculation is performed in an infinite magnetic film,
this type of bounded motion may be relevant for the dynamics
in finite, disc-shaped ferromagnetic elements.

In the special limit $\epsilon\to 0$ the three vortices lie
on a straight line for all times at constant relative distances:
\begin{equation}
C_1 = 2b,\quad C_2=3b,\quad C_3=b.
\end{equation}
Such a linear configuration rotates rigidly around a fixed guiding center
with constant angular frequency
\begin{equation}
\omega = \frac{d\theta_1}{dt} = \frac{d\theta_2}{dt} = \frac{d\theta_3}{dt} = -\frac{1}{6b^2},
\end{equation}
a relation obtained by applying Eqs.~\eqref{eq:eqshamiltonpolarv30} to the present degenerate case.

\section{Conclusions}

We have studied the equations of motion for magnetic vortices which are approximated
as point vortices.
These are a generalization of the Helmholtz-Kirchhoff equations \cite{helmholtz,kirchhoff}
derived for ordinary fluid vortices.
They differ from them in that magnetic vortices have an additional characteristic,
their polarity, which enters in the equations.
The main body of the paper is devoted to the case of three magnetic vortices
which constitute an integrable system.
For the integration of this system
we follow the steps of the seminal paper of Gr\"obli \cite{groebli}
modified to include the vortex polarity.

Magnetic vortices are commonly observed in magnetic elements
with typical dimensions of the order of a few micrometers or hundreds of nanometers.
A single vortex spontaneously created at the center of an element
can be driven by a magnetic field or an electrical current
to produce excitations in its vicinity.
It has been observed in various numerical simulations that
such excitations often have the form of a lump of magnetization opposite
to that of the original vortex.
In some cases the lumps are amplified to become vortex-antivortex pairs
with clearly distinct constituent vortices.
Such a process was apparently observed in experiments
where the original vortex was driven by a magnetic field \cite{neudert05, waeyenberge06}
or an electrical current \cite{yamada07}.
The fact that the created vortex-antivortex pair has opposite polarity to the original
vortex should most probably be attributed to the effect of the 
magnetostatic (dipole-dipole) interaction which is the only magnetic interaction
that primarily favors domains of opposite magnetization.
From these examples it follows that the three-vortex configuration, given in 
Eq.~\eqref{eq:threevortices}, on which we have mainly focused in this paper,
is a particularly interesting one for the purposes of understanding
vortex dynamics in magnetic elements.


In view of the above comments it is interesting to compare our results
with simulations of the Landau-Lifshitz equation.
We have used the Landau-Lifshitz equation with an exchange interaction and an on-site
easy-plane anisotropy so that vortices are the relevant excitations of the system.
Some results have been published in Refs.~\cite{komineas08,komineas09}.
For the case of vortex-antivortex pairs where the vortices are well-separated
the simulations show very good agreement with the analytical solutions for the
vortex trajectories.
It is worth discussing separately the special case of three-vortex collapse 
of Subsection IVB which would appear to be specific to the point vortex system.
We have simulated the case of zero angular momentum within the Landau-Lifshitz
equation and have observed vortex trajectories very similar to those
shown in Fig.~\ref{fig:3}. The original vortex and the antivortex are annihilated
when the three-vortex system collapses to the origin and an isolated vortex 1
is the final outcome of the simulation.
The results of the simulation are closer to the analytical solutions when the
vortex-antivortex pair is large, i.e., the vortex and the antivortex are initially well separated.

\section*{Aknowledgements}
We are grateful to Mahir Hussein for stimulating discussions
that refocused our interest on this subject.


\end{document}